\documentclass[twocolumn,showpacs,aps,nofootinbib]{revtex4}

\usepackage{graphicx}
\usepackage{dcolumn}
\usepackage{bm}

\begin{document}

\preprint{}

\title{
Potential inversion with 
subbarrier fusion data revisited} 

\author{K. Hagino}
\affiliation{
Department of Physics, Tohoku University, Sendai 980-8578, Japan}

\author{Y. Watanabe}
\affiliation{
Department of Physics, Tohoku University, Sendai 980-8578, Japan}

\date{\today}

\begin{abstract}
We invert experimental data for heavy-ion fusion reactions at energies
well below the Coulomb barrier in order to directly determine the internucleus
potential between the colliding nuclei. In contrast to the previous
applications of the inversion formula, we explicitly take into account
the effect of channel couplings on fusion reactions, by assuming 
that fusion cross sections at deep subbarrier energies are
governed by the lowest barrier in the barrier distribution. 
We apply this procedure to the $^{16}$O +$^{144}$Sm and 
$^{16}$O +$^{208}$Pb reactions, and find that the inverted 
internucleus potential are much thicker than phenomenological
potentials. A relation to the steep fall-off phenomenon of fusion cross
sections recently found at deep subbarrier energies is also
discussed. 
\end{abstract}


\maketitle

Nuclear reactions are primarily governed by 
the nucleus-nucleus potential. 
In particular, the Coulomb barrier, 
which appears due to a strong cancellation between 
the attractive nuclear force
and the long-range repulsive Coulomb interaction, 
plays a decisive role in heavy-ion collisions. Several methods have been 
proposed to compute the real part of the internuclear potential. 
Among them, the double folding model has been 
often employed and has enjoyed a success 
in describing elastic and inelastic scattering for many systems
\cite{SL79,BS97,KS00}. 
The Woods-Saxon form has also often been used to parametrize the
inter-nuclear potential \cite{BW91}. 
The surface region of the double folding potential can in fact be well
parametrized by the Woods-Saxon form 
with the diffuseness parameter of around 0.63 fm, and such
phenomenological potential 
has been as successful as the double
folding potential. 

In recent years, many experimental evidences have accumulated that
show that the double folding potential fails to account for
the {\it fusion} cross sections at energies close to the Coulomb 
barrier. That is, 
the double folding potential (and also 
the Woods-Saxon potential which fits elastic 
scattering) overestimates fusion cross sections at energies both 
above and below the Coulomb barrier, having an inconsistent energy 
dependence to the experimental fusion excitation function 
\cite{L95,NBD04,NMD01,HDG02,HRD03,GHDN04,DHNH04}. 
This trend is in accordance with the more recent measurements of fusion cross 
sections at extreme subbarier energies, that show a much 
steeper fusion excitation functions as compared with theoretical
predictions \cite{Jiang}. 


Notice that the scattering process 
is sensitive mainly to the surface region of the nuclear potential, 
while the fusion reaction is also relatively sensitive to the inner part. 
The double folding potential and the Woods-Saxon potential are 
reasonable in the surface region \cite{WHD06}. 
However, 
it is not obvious whether they 
provide reasonable parametrizations 
inside the Coulomb barrier, where the 
colliding nuclei significantly overlap with each other
\cite{NBD04,DP03,ME06}. 
This is so, particularly because the double folding potential takes
into account only the so called knock-on exchange effect, ignoring 
all the other exchange effects originating from the
anti-symmetrization of the total wave function of the colliding
system \cite{AH82}. 

The purpose of this paper is to 
investigate the radial shape of the internucleus potential 
inside the Coulomb barrier and 
discuss its deviation from the conventional parametrizations. 
To this end, 
we apply the inversion formula based on the WKB
approximation \cite{CG78} and 
determine 
the internuclear potential directly from the
experimental data without assuming any parametrization. 
This method was used many years ago 
by Balantekin {\it et al.} \cite{BKN83}. They assumed a
one-dimensional energy independent local potential, and found that the
inversion procedure leads to an
unphysical multi-valued potential for heavy systems. This
analysis has provided a clear evidence for inadequacy of
the one-dimensional barrier passing model for heavy-ion fusion reactions, and
has triggered to develop the coupled-channels approach. 

Although the analysis of 
Balantekin {\it et al.} is important as it has clarified the dynamics
of subbarrier fusion reactions, it is not satisfactory from the point
of view of determination of the internucleus potential. 
Since the experimental evidences for inadequacy of
the double folding potential inside the Coulomb barrier is
increasingly accumulating, it is intriguing to revisit this problem 
by taking into account the current understanding of subbarrier
fusion. 
In this connection, we mention that 
the main reason why 
Balantekin {\it et al.} obtained the unphysical internucleus potentials is 
that they did not take into account the channel coupling effect, 
which has by now been well understood in terms of 
barrier distribution \cite{L95,RSS91,DHRS98,BT98,HB04}. 
Our idea here is to apply the inversion procedure only to the lowest
barrier in the barrier distribution assuming that the fusion cross
sections are determined only by it at deep subbarrier energies. 
We will demonstrate below that the internucleus potentials thus
obtained are well behaved and show a significant deviation from the
conventional Woods-Saxon shape. 

For a single channel system with a potential $V(r)$, 
the inversion formula relates the thickness of the potential, 
{\it i.e.,} the distance between the two classical turning points at a
given energy $E$, with the classical action $S$ as \cite{CG78,BKN83}
\begin{eqnarray}
t(E)&\equiv&r_2(E)-r_1(E) \\
&=&-\frac{2}{\pi}\sqrt{\frac{\hbar^2}{2\mu}}
\,\int^{V_b}_EdE'\,\frac{\left(\frac{dS}{dE'}\right)}{\sqrt{E'-E}},
\end{eqnarray}
where $\mu$ is the reduced mass between the colliding nuclei and $V_b$
is the height of the potential. The classical action $S(E)$ is given
by 
\begin{equation}
S(E)=\int^{r_2(E)}_{r_1(E)}dr\,\sqrt{
\frac{2\mu}{\hbar^2}(V(r)-E)}, 
\end{equation}
and can be obtained once the penetrability $P(E)$ is found in some
way using the WKB relation $P(E)=1/[1+e^{2S(E)}]$. 

In heavy-ion fusion reactions, 
it is well known that the $s$-wave penetrability for the Coulomb
barrier can be approximately obtained from the fusion cross section 
$\sigma_{\rm fus}$ as \cite{BKN83,RSS91,DHRS98,BT98}
\begin{equation}
P(E)=\frac{d}{dE}\,\left(\frac{E\sigma_{\rm fus}}{\pi R^2}\right), 
\label{1stderivative}
\end{equation}
where the effective moment of inertia $R$ may depend on energy
\cite{BKN83,BDK96}. 
This formula assumes that the number of classical turning point 
is two for all partial waves, and thus implicitly assumes 
a deep internuclear potential. 
In the previous application of the inversion formula by Balantekin
{\it et al.}, they assumed that the penetrability so obtained was
resulted from the penetration of a one dimensional energy independent 
potential
\cite{BKN83}. 
Instead, here we assume 
that the penetrability $P$ is given as a weighted sum of contribution from 
many distributed barriers, where the distribution arises due to a coupling 
of the relative motion between the colliding nuclei to nuclear intrinsic
degrees of freedoms such as collective vibrational or rotational
excitations. In this eigen channel picture, the penetrability is given by, 
\begin{equation}
P(E)=\sum_nw_n P_n(E),
\end{equation}
where $P_n$ is the penetrability for the $n$-th eigen-barrier and 
$w_n$ is the corresponding weight factor. 
This concept has been well established by now from the experimental
measurements for the barrier distribution \cite{L95,DHRS98} as well as
from numerical calculations of coupled-channels equations
\cite{HB04,HTB97}. In principle, the weight factors $w_n$ depend on
energy if the excitation energy for the intrinsic motion is not zero. 
However, the energy dependence is shown to be weak\cite{HTB97}, and 
we assume in this paper 
that the weight factors are energy independent.  

At energies below the lowest eigen barrier (i.e., the adiabatic
barrier) in the barrier distribution, one 
expects that only the lowest barrier contributes to the total 
penetrability, 
\begin{equation}
P(E)\approx w_0 P_0(E). 
\label{adiabatic}
\end{equation}
This indicates that one can apply the inversion formula to 
the lowest eigen barrier using fusion cross sections at 
deep subbarrier energies, after correcting the weight factor. 
In order to demonstrate how this works, Figs. 1(a) and 1(b) show the 
second and the first derivatives of the measured 
$E\sigma_{\rm fus}$ \cite{L95} for the $^{16}$O+$^{144}$Sm reaction, 
respectively. 
The former quantity is usually referred to as the fusion barrier
distribution \cite{L95,DHRS98}. We use the point difference formula
with $\Delta E_{\rm c.m.}$=1.8 MeV to carry out the derivatives.
For this system, one can clearly recognize that the barrier
distribution has a double peaked structure. Correspondingly, the first
derivative $d(E\sigma_{\rm fus})/dE$ appears to have two steps as 
a function of energy. 
If one neglects weak couplings to the double phonon state in the
$^{144}$Sm nucleus \cite{HTK97}, the double peaked structure of 
the barrier distribution can be interpreted to originate mainly from
the coupling to the 
first 3$^-$ state in $^{144}$Sm. 
Notice that in general a barrier distribution has a peak at 
the energy equal to the height of a potential barrier. 
Assuming that the main peak of the barrier distribution around 
$E_{\rm c.m.}\sim $60 MeV consists only 
of the contribution from the lowest eigen barrier, we 
scale the first derivative $d(E\sigma_{\rm fus})/dE$ so that it has a
value of 0.5 at the peak energy, which we assume to be identical to
the position of the lowest barrier, $V_b$.  
The scaling factor corresponds to the product of the 
geometrical factor $\pi R^2$ and the weight factor $w_0$ 
(see Eqs. (\ref{1stderivative}) and (\ref{adiabatic})). 
The function thus obtained is shown by the filled circles in
Fig. 1(c). This function can be interpreted as the penetrability for 
the lowest barrier, to which one can apply the inversion formula to
determine the radial shape. 

\begin{figure}[htb]
\includegraphics[scale=0.5,clip]{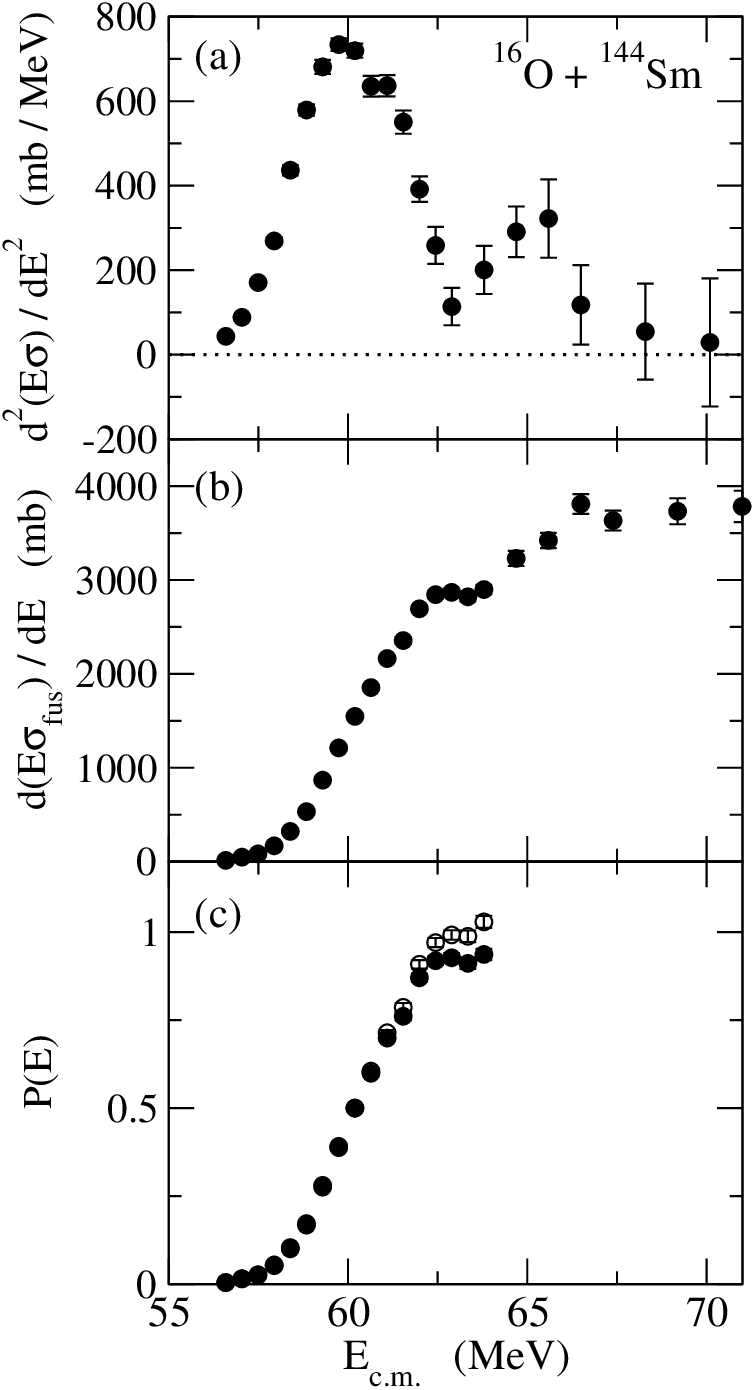}
\caption{
Fig. 1(a): The experimental fusion barrier distribution 
for the $^{16}$O+$^{144}$Sm reaction defined as 
$d^2(E\sigma_{\rm fus})/dE^2$. 
The experimental data are taken from Ref. \cite{L95}. 
Fig. 1(b): The first derivative of 
$E\sigma_{\rm fus}$ 
for the $^{16}$O+$^{144}$Sm reaction. 
Fig. 1(c): The same as Fig.1 (b), but normalized so that it is 0.5 at
the energy of the lower peak in the barrier distribution shown in
Fig. 1(a) (the filled circle). The open circles are obtained in a
similar way, but by taking into account the energy dependence of the
effective moment of inertia as given by Eq. (\ref{inertia}). 
}
\end{figure}

The inversion formula yields only the barrier thickness, $t(E)$, and
one has to supplement either the outer or the inner turning points 
to determine the radial shape of the potential \cite{BKN83}. 
We estimate the {\it outer} turning point $r_2(E)$ using the 
Coulomb interaction of point charge and the 
Woods-Saxon nuclear potential, 
\begin{equation}
V_N(r)=-\frac{V_0}{1+\exp[(r-R_0)/a]},
\label{WS}
\end{equation} 
with the range parameter of 
\begin{equation}
R_0=\sum_{i=P, T} 
\left(1.233A_i^{1/3}-0.98A_i^{-1/3}\right)+0.29 ~~~~({\rm fm}), 
\end{equation}
and the diffuseness parameter of $a$=0.63 fm. 
We adjust the depth $V_0$ in order to reproduce the barrier height
$V_b$ determined from the peak position of the barrier distribution. 
Since the Coulomb term dominates at the outer turning point, 
except for the region near the barrier top, 
the inverted potential is insensitive to the actual shape of nuclear 
potential employed to estimate the outer turning point. 
The Woods-Saxon potential (\ref{WS}) determines not only the outer
turning point but also the position of the potential barrier, $R_b$. 
Following Ref. \cite{BKN83}, we use 
\begin{equation}
R(E)=\frac{1}{2}\,\left(R_b+Z_PZ_Te^2/E\right), 
\label{inertia}
\end{equation}
for the effective moment of inertia in Eq. (\ref{1stderivative}). 
The penetrability obtained by taking into account the energy
dependence of the effective moment of inertia is denoted by the 
open circles in Fig. 1(c). Although the difference between the filled
and open circles is not large, especially at energies below the
barrier, the penetrability behaves slightly better if one considers
the energy dependence of moment of inertia, since it is saturated 
at unity at high energies. 
In the actual application of the inversion formula shown below, we 
smooth the data points with a
fifth-order polynomial fit to the function 
$\ln[E\sigma_{\rm fus}/\pi R(E)^2]$ \cite{BKN83}. 
We have confirmed that the results do not significantly change even if
we use a higher order polynomial fit. 
We also fit the lowest peak of the barrier distribution using the Wong
formula \cite{W73} in order to accurately estimate the barrier height $V_b$. 

We have confirmed the accuracy of the inversion procedure using the 
theoretical fusion cross sections obtained by the computer code {\tt CCFULL}
\cite{HRK99}. For this purpose, we consider the 
$^{16}$O+$^{144}$Sm system, and generate the fusion cross sections by 
taking into account the excitation to the first 3$^-$ state in 
$^{144}$Sm. We use the same parameters as in Ref. \cite{HTD97}. 
We find that the resultant inverted potential closely follows the 
adiabatic potential obtained by diagonalising the coupling 
Hamiltonian at each position $r$ \cite{HB04,HTB97}. 
This evidently justifies our procedure for the potential inversion discussed
above. 

\begin{figure}[htb]
\includegraphics[scale=0.5,clip]{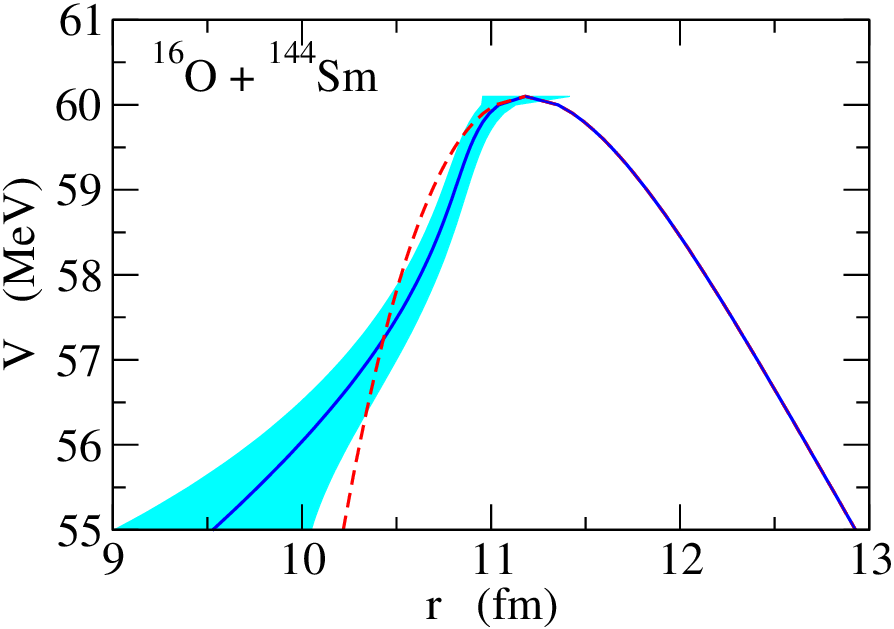}
\caption{
(Color Online) The adiabatic potential for the 
$^{16}$O+$^{144}$Sm reaction obtained with the inversion method. 
The dashed line is a barrier due to a phenomenological Woods-Saxon potential. } 
\end{figure}

We now invert the experimental data for 
the $^{16}$O+$^{144}$Sm system shown by the open circles in Fig. 1(c)
in order to obtain the radial dependence of the adiabatic barrier. 
The result of the inversion method is shown in Fig. 2. 
The uncertainty of the inverted potential is estimated in the same way
as in Ref. \cite{BKN83}. 
The dashed line shows the barrier due to the Woods-Saxon potential (\ref{WS}) used to
estimate the outer turning points. 
One clearly sees that the inverted potential is much thicker than the 
phenomenological potential at low energies, although 
it is close to the phenomenological potential at energies close to 
the potential barrier. 
This trend is opposite to what Balantekin {\it et al.} found
in the previous analysis. If there was an unresolved peak in the
barrier distribution below the main peak, one would obtain a much
thinner barrier than the phenomenological potential, as in the previous 
analysis. We have actually obtained 
such unphysical thin barriers for the $^{17}$O + $^{144}$Sm and 
$^{16}$O + $^{148}$Sm systems, where the main peak of the barrier
distribution is not expected to correspond to the lowest eigen barrier
\cite{L95}. Having a thick barrier, rather than a thin barrier, we are 
convinced that the main peak of the barrier distribution for 
the $^{16}$O+$^{144}$Sm reaction indeed consists of the lowest eigen
barrier. In this way, the potential inversion method could also be used
to judge whether there is an unresolved barrier in the barrier
distribution. 

\begin{figure}[htb]
\includegraphics[scale=0.5,clip]{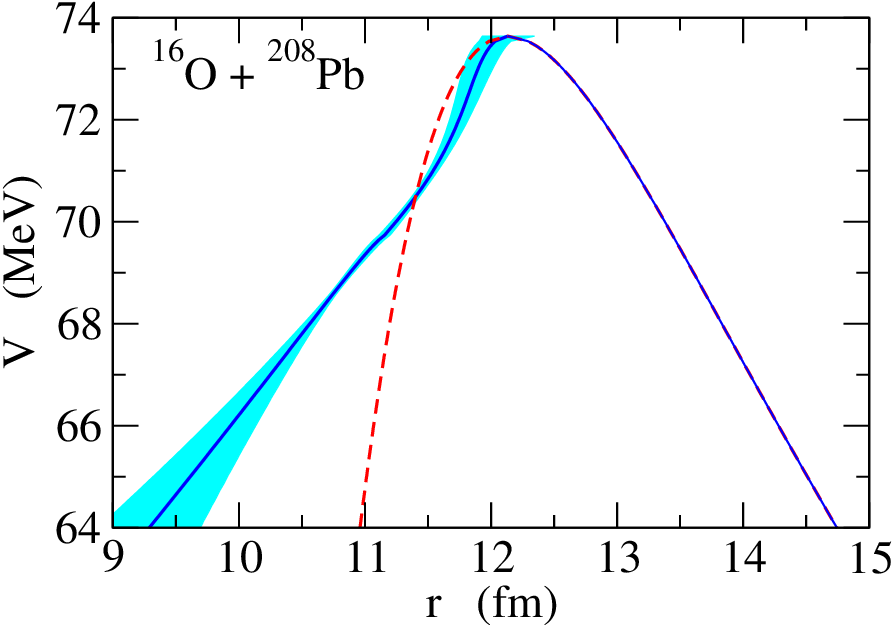}
\caption{
(Color Online) Same as Fig. 2, but for the 
$^{16}$O+$^{208}$Pb reaction. }
\end{figure}

For the $^{16}$O+$^{144}$Sm reaction, the experimental data exist only
from $E_{\rm c.m.}$=56.6 MeV (that is, about 3.6 MeV below the lowest
barrier height), and the inverted potential has a large uncertainty 
below this energy. In order to discuss the behaviour of the potential
far below the barrier, we next consider the 
$^{16}$O+$^{208}$Pb reaction, for which the fusion cross sections were
recently measured at deep subbbarier energies \cite{DHLN06}. 
Figure 3 shows the potential barrier for this system obtained with the
inversion method. 
One sees that 
the behaviour of the potential is qualitatively similar to that for the 
$^{16}$O+$^{144}$Sm reaction shown in Fig. 2. 
Namely, the inverted potential is close to a phenomenological 
potential in the region near the barrier top, but  
it deviates largely at deep subbarrier energies, where 
the thickness is much larger than the phenomenological potential. 
The thicker the potential is, the smaller the penetrability is, and
also the stronger the energy dependence of the penetrability is. 
The thick potential barrier obtained for the 
$^{16}$O+$^{144}$Sm and $^{16}$O+$^{208}$Pb systems is thus consistent 
with the recent experimental observations  
\cite{Jiang,DHLN06} that the fusion excitation
function is much steeper than theoretical predictions at deep
subbarrier energies. Although the present analysis does not exclude a
possibility of a shallow potential \cite{ME06}, the present study
suggests that the origin of the
steep fall-off phenomenon of fusion cross section can be 
at least partly attributed to the departure of internuclear potential 
from the Woods-Saxon shape. 

For the $^{16}$O+$^{208}$Pb system shown in Fig. 3, the deviation of
the inverted potential from the phenomenological potential 
starts to occur at around $E=70.4$ MeV. 
It is interesting to notice that this energy is very close to 
the potential energy at the contact configuration estimated with the 
Krappe-Nix-Sierk potential\cite{KNS79,IHI07}. 
Inside the touching configuration, the potential 
represents 
the fission-like adiabatic potential energy surface. 
The effect of such one-body potential has been considered recently
and is shown to account well for the steep fall-off phenomena of
fusion cross sections \cite{IHI07}. 
The inverted potentials which we obtain are thus intimately related to
the one-body dynamics for deep subbarrier fusion reactions. 

In summary, we applied the potential inversion method, which relates
the potential penetrability to the thickness of the potential barrier, 
in order to investigate the radial dependence of the internucleus
potential for heavy-ion fusion reactions. 
To this end, we assumed that the tunneling is well described by the
lowest adiabatic barrier at deep subbarier energies, and extracted the
penetrability by combining the experimental barrier distribution and 
fusion cross sections. 
We found that the resultant potential for the 
$^{16}$O+$^{144}$Sm and $^{16}$O+$^{208}$Pb systems 
is much thicker than a barrier obtained with 
a phenomenological Woods-Saxon potential. 
This indicates that the steep fall-off phenomenon of fusion cross
sections recently observed in several systems can be partly accounted
for in terms of a deviation of internuclear potential from the 
Woods-Saxon shape. 

The lowest peak in a barrier distribution is relatively well resolved in
general for systems involved with a vibrational target. It would be an
interesting future work to apply the inversion method systematically
to such systems and discuss a global potential for heavy-ion
collisions. Another problem is the dynamical effects after the
touching configuration. 
In this paper, we exploited a barrier distribution picture for
subbarrier fusion, but assumed an energy independent potential for 
each distributed potential barrier. 
It would be an interesting work to discuss how the
thick potential which we obtained in this paper 
is related to dynamical effects such
as the coordinate dependence of reduced mass, and 
energy and angular momentum dissipations, that are other promising 
origins for the steep fall-off phenomenon of fusion cross 
sections \cite{NMD01,DHNH04,DHLN06}. 

\bigskip

We thank N. Takigawa, M. Dasgupta, and D.J. Hinde 
for useful discussions. We thank M. Dasgupta also 
for providing us with the experimental data prior to
publication. 
This work was supported by the Grant-in-Aid for Scientific Research,
Contract No. 19740115, 
from the Japanese Ministry of Education, Culture, Sports, Science and
Technology.

\end{document}